# A Wikipedia Literature Review

Owen S. Martin

May 12, 2010

## 1 Introduction

Wikipedia is quickly becoming one of the most complex and most relevant data sets humanity has ever produced. While some efforts exist to deny its credibility, it is rapidly stepping in to being the starting point for anyone's intellectual curiosities. The Wikimedia Foundation early on recognized the need for all sorts of revision protection and user communication to ensure the quality of such a novel project. The result is an Encyclopedia with over three million articles, ten million registered users, and a staggering 368 million total number of edits.[1]

Indeed, the robustness of the Wikipedia system lies fundamentally in the fact that every single edit to every page is stored permanently on Wikipedia's servers. Thus, not unlike with version control in software development, no single edit can destroy the existing information. However, with 368 million edits available, knowing the exact state of Wikipedia at any given time is almost akin to knowing the positions and velocities of every molecule in a tank of air: research has persuaded scientists to model the overall behavior as a probabilistic phenomenon rather than a deterministic one. Similarly with Wikipedia: especially given that the data within Wikipedia is unstructured text, we are reaching a state where no computer scientist can write a program which can scale to the entirety of the data.[2]

The need to introduce statistical inference to the problem is evident. Not only has software become unscalable for this task, we recognize that the growth pattern of Wikipedia is essentially random, not unlike any ecology or economy. This paper does not propose to state and solve the Wikipedia problem *in toto*, but rather give insight into how existing statistical methods in anomaly detection [and machine learning, trust, etc.] can be brought to bear on such a problem, and how the novel structure of Wikipedia can motivate innovations in those technologies.

---

[1] At the time of this writing.
[2] Even Wikipedia's servers are now unable to dump all of the data into a *highly* compressed file of about 255 GB in a reasonable amount of time. The last 6-month attempt failed towards the end due to propagating errors.

This literature review will come in several discrete steps. First, the nature of Wikipedia must be described and catalogued, so that mathematical rigor and statistical integrity can be applied to it. Second, the history of the existing statistical technologies will be presented, for reference to the changes we apply for this problem.

# Introduction

Wikipedia, the free online encyclopedia, was launched in January 2001 by Mr. Jimmy Wales and Dr. Larry Sanger as an alternative to the now-defunct Nupedia, which was to be a free, collaborative encyclopedia written exclusively by experts and reviewed by peers Olleros (2008). It is based on Ward Cunningham's Wiki Wiki (derived from the Hawaiian phrase *wiki wiki* meaning "quick"), a software that allowed for open contributions from any registered or anonymous users. This software allows any user to modify any page at any time. It saves every previous version of a page and makes all page histories available to all users. Thus a user can revert any errors or graffiti left by his predecessor using the built-in *revert* tool.

# 2 The Relational Database

Wikipedia evidently constitutes an extremely large data set and for fast retrieval is stored as a relational database. The schema for this database is available on meta-Wikipedia pages, which are Wikipedia pages about Wikipedia. A graphical representation found at Manual:Database layout. Furthermore, precise definitions for each of the tables to be described in this section can be found at Category:MediaWiki database tables.

The database contains more information that what will be presented here, but this summary aims to categorize all the information available for mining the internal Wikipedia structure. Omitted information includes caches for expensive computations for the database, such as grouped queries, as well as parser testing and other server issues.

Mediawiki provides four central tables to the database which supply foreign keys to other tables. Somewhat intuitively, these are the `page`, `text`, `user`, and `image` tables. We will survey the structure of the first three of these tables, recognizing that mining images is beyond the scope of this summary.

These three tables provide foreign keys to peripheral tables which add essential structure to Wikipedia, including categorization, navigation, and other forms of communication. It is this added structure that makes statistical data mining feasible and interesting. Note that hyperlinks can be found throughout this work for direct reference to the respective Mediawiki manual pages.

## 2.1 The `page`

The page table is said to be the "core of the wiki". It contains a unique primary key `page_id`, as well as some essential metadata. The page table contains the revision id for the latest revision, its length in bytes, and a unique page title, referred to as a *sanitized* page title as it uses capitalizations, underscores, and punctuation marks consistently, which is often referenced elsewhere in the database instead of the primary key. It also contains a counter for the number of times the page has been visited, but for Wikipedia and other Mediawiki sites this feature is disabled. Fortunately, this data is available for Wikipedia via `http://stats.grok.se/` as it is essential for inquiries into site traffic and estimating page relevance.

The *namespace* of the page enables one to distinguish the type of metadata that appears on that page. The namespace is essential to understanding the structure of Wikipedia.

### 2.1.1 Namespaces

Namespaces appear at the end of a Wikipedia URL as a prefix to the page name, followed by a colon. An example appears on the manual page:

| Title | Namespace |
|---|---|
| `Foo` | Main |
| `Project:Foo` | Project |
| `Help:Foo` | Help |

So the Main namespace requires no prefix. Namespaces are used to distinguish what type of metadata appears on the page. The Main namespace is of course primary and includes the articles themselves. Other namespaces including talk, help, and user are found in Table 1.

Wikipedia is an instance of the MediaWiki software, and thus has its own customized namespaces specific to the project. See Table 2. They are essential to Wikipedia structure by separating data from metadata. A prime example is to consider the difference between the pages

- `http://en.wikipedia.org/wiki/Wikipedia`

| Index | Name | Purpose |
| --- | --- | --- |
| 0 | Main | "Real" content; articles |
| 1 | Talk | Talk pages of "Real" content |
| 2 | User | User pages |
| 3 | User talk | Talk pages for User Pages |
| 4 | Project | Information about the wiki |
| 5 | Project talk | |
| 6 | File | Media description pages |
| 7 | File talk | |
| 8 | MediaWiki | Site interface customisation |
| 9 | MediaWiki talk | |
| 10 | Template | Template pages |
| 11 | Template talk | |
| 12 | Help | Help pages |
| 13 | Help talk | |
| 14 | Category | Category description pages |
| 15 | Category talk | |
| -1 | Special | Holds special pages |
| -2 | Media | Alias for direct links to media files |

Table 1: The Index, Name, and Purpose of built-in Mediawiki namespaces. Negative indices are assigned to namespaces that are not regularly accessed, but exist for background processing purposes.

- `http://en.wikipedia.org/wiki/Wikipedia:About`.

The former article falls in namespace 0, an article. The latter falls in namespace 4, the project. Thus for Wikipedia the prefix `Project:` has the alias `Wikipedia:`, hence they point to the same page. Wikipedia further shortens many namespaces, the most common being `WP:` for `Wikipedia:`. Note, however, that the aforementioned *sanitized* page title will be the one with the `Wikipedia:` prefix, and other prefixes, including `Project:`, will redirect the user to the sanitized URL.

### 2.1.2 Templates

Templates are a prevalent feature of Wikipedia and appear on most developed pages. In the wiki markup language, a template is called using

$$\{\{\texttt{Template:}template\_name|parameters\}\}.$$

| Basic namespaces | | Talk namespaces | |
| --- | --- | --- | --- |
| 0 | Main | 1 | Talk |
| 2 | User | 3 | User talk |
| 4 | Wikipedia | 5 | Wikipedia talk |
| 6 | File | 7 | File talk |
| 8 | MediaWiki | 9 | MediaWiki talk |
| 10 | Template | 11 | Template talk |
| 12 | Help | 13 | Help talk |
| 14 | Category | 15 | Category talk |
| 100 | Portal | 101 | Portal talk |
| 108 | Book | 109 | Book talk |
| Virtual namespaces | | | |
| -1 | Special | | |
| -2 | Media | | |

Table 2: The Wikipedia Namespaces. As in table 1 the virtual namespaces are rarely accessed and are used for background processing.

Templates, as the name suggests, allow the reuse of certain formatting efforts. This is achieved using transclusion, whereby an entire page is called and reused within the calling page. The prime example is the Infobox, which generates a box with standard information for that type of page. For an article on a city, for example, the infobox would contain the population, mayor, land area, and so forth. These infoboxes, since they are in standard form across articles, are used for data mining and semantic extraction (Wu and Weld, 2007).

Templates are also quite generalizable, and in fact many templates call on other templates. There are 96 of these metatemplates at the time of this writing, each one protected because they are essential to the existence of so many other pages.

Virtually any developed page will employ some template usage, such as a table of contents, an infobox, or subject membership on an article page; project membership or personal information on a user page; or even embedding an image on any kind of page. It is somewhat clear that the ability to repackage formatting work through templates has aided in Wikipedia's growth.

The database contains a list of all the templates used for a particular page specified by a foreign key. The templates are given as a sanitized name with prefix `Template:` and are all found in namespace 10 as indicated in Table 2.

### 2.1.3 Categories

Wikipedia contains an expansive hierarchy of categories. These categories can be found in both articles and other namespaces, although such categories do no cross in and out of the main namespace. Wikipedia pages are formatted to list the categories a page belongs to at the bottom of the page.

Categories are found in their own namespace, `Category:` and pages within this namespace contain a summary of the category, a list of all subcategories (including the number of pages within each subcategory and the number of subsubcategories), a list of all the pages that are found in that category, and a list of all the images found on the pages in that category. A list of *super*categories is found at the bottom of the page as the list of categories this category belongs to.

A simple example of a hierarchy of categories is

$$\{\text{Statistics}\} \supset \{\text{Probability Distributions}\} \supset \{\text{Continuous Probability Distributions}\}.$$

The database contains a table which maps the `page_id` to a category. Thus, the categories any page belongs to can be fetched.

### 2.1.4 Wikilinks

The database will also fetch the internal *pagelinks*, also known as *wikilinks*, for any given page. The `page_id` is the foreign key to this table and the other variates are the namespace and sanitized page name.

Internal wikilinks provide not just a method for a user to browse or surf Wikipedia, they also give an insight into the structure and relevance of existing pages. An obvious heuristic for what constitutes an *important* page is one that has relatively many other pages linking to it. This does have faults to it though: first, virtually any word on a page can be wikilinked, whether it's relevant or not. For example, for the Normal Distribution article, one can wikilink the phrase "United States", which appears in the second paragraph by putting [[ ]] around it in the wiki markup. One would not want to associate a high relevance between the Normal Distribution and the United States. There does exist an effort among Wikipedians to remove irrelevant links like these, but one must accept that there be a few misleading wikilinks scattered throughout the project.

A second drawback to the wikilinks table is that a wikilink is only noted one time for its existence on a page. Thus, for the Normal Distribution page, all we know is that there is at least one wikilink to Carl Friedrich Gauss and at least one wikilink to Confluent hypergeometric functions, but one is clearly more closely related.

Using the wikilink table, an abstract graph can be synthesized of the entire Wikipedia, or some designated subsection (for example, the category "Statistics"). The usual network analysis has been performed on this graph (Warren et al., 2008; Voss, 2005, e.g.). Graphs that are generated from a random or unknown process are often categorized into two types: the Erdös-Rényi random graph (Erdös and Rényi, 1959), in which the probability of an edge existing is basically constant across all edges of the complete graph, or the so-called "scale-free" graph, in which edge distribution follows a power law (Barabasi, 2003). The term *power law* is shorthand for any variation on Zipf's law (Zipf, 1949) or the Pareto distribution (Pareto, 1901). Wikipedia's wikilink graph is generally found to be scale-free, as Barabasi (2003) would expect given that it is generated from a social networking process.

### 2.1.5 Page Restrictions

One of the most controversial topics regarding Wikipedia is restricting users from editing certain pages. The Page Restrictions table contains all restrictions that have been put in place, using its own primary key (`pr_id`)and calling on `page_id` as a foreign key. The `pr_type` is the type of protection (whether it applies to edits, page moves, or similar), and the `pr_level` describes the level of protection for the page; full protection for pages only editable by sysops, semi-protection for autoconfirmed users, or other levels including move-protection, creation protection, and upload protection.

The `pr_cascade` field determines whether cascading protection is implemented. This would imply that all transcluded templates and images on the page will be protected as well. Transclusion, as mentioned above, involves reusing the content on a particular page in another page.

### 2.1.6 Language Links

Many pages of Wikipedia exist in other languages. The English Wikipedia is far and away the largest one, but most salient articles will appear in many of the world's commonly spoken languages. For any one page, one can extract the other languages' Wikipedias which contain the same article and the name of that article in the other language.

### 2.1.7 External Links

Wikipedia pages often feature links to webpages outside of Wikipedia. These links are shown in a different hyperlink color and are marked with an external link dingbat.

It is possible to query for all the external links coming from a particular Wikipedia page.

Wikipedia has somewhat strict regulations on what type of external links are allowed to appear on a page. They can be found at WP:EL.

### 2.1.8 Redirects

Some pages in Wikipedia serve only as *redirects*. That is, they immediately send the user to another page. This is used for handling multiple titles for the same article. For example, "Normal Distribution" is redirected to "Normal distribution", since Wikipedia pages are case sensitive. Similarly "Gaussian distribution" is redirected to "Normal distribution", as they are one in the same.

The class of redirect pages is essential for what is meant by a *sanitized* Wikipedia title above. Indeed when a table elsewhere in the database refers to a page by its title, it should reference the actual article and not a redirect. While this used to be fixed by hand, by now these references are almost always fixed by bots (see below for more details on bots).

### 2.1.9 Image Links

Finally, many Wikipedia pages feature images stored on the Wikimedia Commons database. (The Wikimedia Commons should not be confused with *Mediawiki*, which is the software engine behind Wikipedia.) A table exists in the Wikipedia database of links from Wikipedia pages to the location on the Wikimedia Commons database where the images are stored.

## 2.2 The `user`

The `user` is the second central table mentioned in this survey, and provides a master list of all registered users using a primary key `user_id`. The table contains basic account information for each user, including their username, "real name", password, preferences, and email address. The primary key that appears here acts as a foreign key in many other essential tables.

### 2.2.1 Watchlists

A registered user can assign particular pages in any namespace to her own watchlist, which contains the `user_id` as a foreign key, the sanitized article title as the page being watched, and a timestamp for generating change notifications. Thus, a user with an interest in a page can be kept up to speed with automatic notifications from Wikipedia on any changes made to the page. This is one of the devices through which Wikipedians effectively combat vandalism, and one of the main forms of communication among Wikipedians.

### 2.2.2 User Groups

The user groups table stores the memberships of each user to the various user access levels available. Wikipedia has a particular instance of these access levels, found at WP:UAL.

When an account is created, that user is automatically assigned from the regular user levels: Anonymous users, New users, Autoconfirmed users, and Confirmed users. Thus any anonymous user is still given a user i.d., a user name which is their I.P. Address, and assigned to the group Anonymous user. Users who register are assigned to the New user group, and given a threshold of legitimate Wikipedia activity, monitored by Wikipedia's web-based machines called *bots*, described below, are assigned into the Autoconfirmed group. Once Autoconfirmed, a user can move pages, edit semi-protected pages, and upload images.

Higher-privileged groups include *administrators*, *bureaucrats* and *stewards*. The existence and necessity of the administrator role is a large source of controversy for Wikipedia (as is any socio-economic inequality of privilege), but administrators, also colloquially known as *sysops*, carry out many essential functions. The administrator's tools cover processes such as page deletion, page protection, blocking and unblocking, access to modify fully protected pages and the mediawiki interface, and the ability to grant and remove *account creator*, *rollback*, *ipblock-exempt*, *confirmed user*, *autoreviewer*, and *edit filter manager* rights to other users.

These latter rights somewhat intuitively follow their respective names, but begin to go beyond the scope of this summary. Details are available at WP:UAL.

For simplicity, bureaucrats are administrators with the additional ability to add users to the sysop and bureaucrat groups, as well as the ability to add and remove users from the *bot* group. Technically one can be a bureaucrat without being a sysop, but no instance of this has occurred. Stewards are bureaucrats with the ability to revoke sysop and bureaucrat status, among other permissions.

Finally, the *bot* user group plays a special role for researchers of Wikipedia, as it is the user class of non-human automata that can gather information and make necessary edits faster than any human user. There is a large class of bots that exist today, and as they are machines, they are not programmed for complex tasks but rather for trivial and repetitive ones, including dead link removal and redirect fixing.

Any bot that exists on Wikipedia is equipped with an emergency shut-down button which must be prominently displayed on the bot's user page. Bots go through a thorough review process before being approved by administrators. Among the general guidelines for bots is that they be harmless and useful.

A researcher, particularly one interested in Statistics, would generally create a non-editing bot: one that goes and retrieves information. While the Wikimedia administration discourages the creation of these types of bots, suggesting that data mining is best performed on a database dump, we have already seen the size of these data dumps render the computing nearly intractable. Furthermore, an effort to avoid hogging Wikimedia bandwidth motivates parsimonious sampling methods and computational providence.

For example, bots can perform simple surveys. They can return a list of all the pages that fall in the category "Continuous Probability Distributions" or all the pages that link to "Chi-square_test" or all the pages that employ the Statistics Navigational Template. They can also check to see if some binary condition is satisfied, for example, whether or not the page is considered a stub, and return the sum of the pages that do satisfy the condition.

### 2.2.3 Discussion Page Notifications

There exists extensive communication across Wikipedia. The three main forms are to edit the discussion page of the article in question, to email a user directly, and most commonly, to edit the discussion page of user. Just as watchlists give the user a summary of all the changes to pages she is interested in, the discussion page notification is a more active "you have new messages" feed.

### 2.2.4 Logging

Wikipedia keeps a rolling log of changes, so that filtered syndication can be delivered to interested Wikipedians. The Logging table includes a primary key of distinct logging events, and uses the `user_id` as a foreign key to indicate which user is responsible for the action. The action type typically falls under: *block, delete, import, makebot, move, newusers, protect, renameuser, rights*, and *upload*. Each type has a collection of actions corresponding to it. Each item in the Logging table includes a timestamp, an indication of namespace, the sanitized title of the affected page, and comments.

The Logging table is therefore highly dynamic and interesting to researchers involved in real-time or highly time-sensitive studies. To sort through all this rolling data, a somewhat standardized tag system has been implemented, Special:Tags.

### 2.2.5 IP Blocking

One of the most salient features of Wikipedia and an essential safeguard against vandalism is the ability of administrators to block users. The `Ipblocks` table contains the users who have been blocked and the user who added the block as a foreign key from the `user` table. Also appearing in the table are the relevant timestamp, a concise reason for the block, and an expiry date on the block. It is also possible to block a range of IP addresses, in the case of a centralized attack, and such a block would appear in this table using `range_start` and `range_end` entries.

## 2.3 The `text`

Finally, arguably the most important part of Wikipedia is stored in a general `text` table. This table is obviously quite extensive as it must contain all of the text that appears anywhere in Wikipedia's history. However, the table contains only three variates: `old_id`, which acts as a primary key, `old_text`, a `mediumblob` containing the text (which can be quite large), and `old_flags`, which is used mainly to indicate whether or not the text is compressed. The *gzip* compression method is essential for storing such an extensive data set.

The `text` table is intentionally kept simple so other more complex tables can refer to it. It provides foreign keys to the `revision`, `recentchanges`, and `archive` tables. Note that it does not provide a foreign key directly to the `page` table, but rather that the page table calls the text of the page *through* the revision table.

### 2.3.1 Revisions

At the time of this writing, there have been over 365 million revisions made to Wikipedia, each of which revision's metadata is stored in the `revision` table. Because the table is so large, it is indexed by a composite key made up of a primary `rev_id` key as well as respective `page_id`, `user_id`, and `text_id` foreign keys. The table also contains a timestamp, which is in fact the only indication of which revision follows which. The length of the revision, in bytes, is also available, which is useful to label an article as "mature" or "evolved." Any article whose magnitude of change is below a percentage can be named; one can model the rate of change.

### 2.3.2 Recent Changes and Archive

The `recentchanges` table contains information about the latest modifications done to the wiki (not older than 30 days). The contents of this table are used to generate the recent changes pages, related changes pages, watchlists, and the list of new pages. While this table is mainly used for back-end processing, a user can visit Special:RecentChanges or join the Recent Changes Patrol at WP:RCP as yet another method of monitoring important changes in real time.

The `archive` table contains metadata on deleted pages. Just as deletions within a page can be reverted on Wikipedia, so can an entire deleted page be revived. The table contains the namespace and article title of the deleted page, and contains a timestamp of the last revision. The actual time of deletion, however, is found in the `logging` table above.

## 2.4 Other features

The Mediawiki software comes with a hitcounter for any page to track usage. This feature has been disabled on Wikipedia for many reasons, not the least of which is the bandwidth that must be made available to include the hitcounter in most queries. However, the Wikipedian User:Henrik has direct access to the usage data and publishes it at grok.se in many user-friendly formats, including JSON. The statistics published are daily and monthly traffic for any Wikipedia page in any namespace, and generally go back to late 2007. This data is essential for any model that proposes to compare Wikipedia changes and structure with Wikipedia use.

# 3 Recent Academic Inquiries

Research directly on Wikipedia is largely absent from the field of Statistics. Computer Scientists, on the other hand, have been presenting results at Artificial Intelligence conferences and Wikipedia symposia for the last five years. They have mainly plucked the low-hanging fruit using statistical, data mining, and machine learning techniques with abandon. As this thesis proposes to introduce novel statistical methods to Wikipedia analysis, it is necessary to survey what work has been done.

## 3.1 Quality

The definition of *quality* across the literature is neither formal nor standardized. Thomas and Sheth (2007) characterize a *good article* as one that is reliable for information extraction, a topic that has seen extensive formal research (Manning et al., 2008, e.g.). However, formal distinctions among quality, trust, reliability, and reputation are not made consistently across the literature.

A gentle introduction to what is generally considered the problem of quality is found in Blumenstock (2008). The author reviews various efforts towards supervised prediction of high-quality articles, using the "featured article" tag for supervision. He chooses a model that uses just word count as a predictor of featured article status, and in fact choosing 1830 words as a cutoff point for featured status achieves 96.46% accuracy. Many other predictors are mentioned in the paper, as found in table 3.

| **Frequency Counts** | | |
|---|---|---|
| character count | complex word count | token count |
| one-syllable word count | sentence count | total syllables |
| **Readability indices** | | |
| Gunning fog index | FORCAST formula | Flesch-Kincaid |
| Coleman-Liau | Automatic Readability | SMOG index |
| **Structural features** | | |
| reference links | reference count | section count |
| internal links | category count | citation count |
| external links | image count | table count |

Table 3: Well-used classification features, from Blumenstock (2008).

The author finds that in a handful of randomized experiments, simple word count classified most accurately, and that the increase in accuracy by adding another one or more of these features to the classifier does not warrant the sacrifice in parsimony. However, table 3 is useful to understand what features generally appear in the literature as easily accessed, important, and predictive.

Using one simple quantitative classifier and simply minimizing the misclassification value on that feature, high classification accuracy was achieved without the use of sophisticated classification such as $k$-means clustering, Bayesian classifiers, and support vector machines. However, this sheds light on the fact that in this case *quality* has an oversimplified definition: whether or not it has a "featured" tag. Furthermore, it does not begin to question causality, even though a researcher would be unlikely to blindly believe that an article being long would imply that it was high-quality.

To dig deeper into the question of quality, one must first investigate the motivations a research would have for a reliable (if not automatic) article quality measure. One can decompose the question of quality into two general categories: *immunity to vandalism* and *level of development*.

### 3.1.1 Vandalism

Potthast et al. (2008) describes three type of delinquents:

(i) *Lobbyists*, who try to push their own (usually political) agenda;

(ii) *Spammers*, who solicit money for products or services; and

(iii) *Vandals*, who deliberately destroy the work of others.

This last category is perhaps the easiest to spot, as the authors point out that usually vandalism consists only of "gobbledygook," such as

> a string of random characters ... if the keyboard is hit randomly, ... capital letters, ... repetition of characters, ... large parts of an article deleted ...

These events might not be as easy for a machine to discover, but can still be classified as features as in table 3. The idea is to program an automatic vandalism detector, which not unlike Blumenstock (2008) and others, classifies Wikipedia pages into the vandalized and non-vandalized.

Once again various classifiers and machine learning techniques can be brought to bare on the question, but one must ask the scientific merit to building such machines. The fact of the matter is that any such machine *must* be trained, at least at first, by human supervision. But when it comes to Wikipedia, human supervision is always available. Even a well-designed and well-trained classification machine would almost never outperform a single human user, much less many human users, nor would we even want it to. Therefore the usefulness of automatically detecting vandalism with machines is negligible, especially for the latter two categories of delinquency: vandalism and spam.

The question of lobbyists, however, goes somewhat deeper. There have been obvious cases of users pushing political agendas, usually in a disorganized fashion, on pages of high controversy, such as George W. Bush or Karl Marx. But there have also been instances of a well-organized and subtle attack on entire subjects, using Sock Puppets and Meat Puppets, for example resulting in the time Wikipedia arbitrators decided to block all IP addresses coming from known Scientology servers (WP:ARSCI).

Lobbyists are not always in the wrong, however. Consider the early 17th century, when Galileo published his theory of heliocentricity (Galilei, 1632). Had Wikipedia existed at the time, his edits would have been immediately reverted. Therefore the question of the scientific method cannot be avoided in Wikipedia. Today's world sees much more research both in science and in liberal arts, and so often the challenging of accepted norms that a governance and citation system is needed in Wikipedia, not unlike the citation system developed for academic articles.

The evolution of a scientific or political theory over time is a generalization of what's known in the literature as *topic drift*. The substantial reduction of change from an entire theory to just the frequency of certain words or $n$-grams appearing on blogs and Wiki pages gives rise to stochastic models of this change over time. Knights et al. (2009) estimate drift by defining indicators of drift such as *perplexity*, which they define as

$$2^{-\sum \frac{1}{N} \log_2 p(\text{word}|\text{model})},$$

or the usual Kullback-Leibler divergence

$$\text{KLD}(P\|Q) = \sum p_i log(p_i/q_i).$$

Once again heuristics and learning machines are used to better quantify what is meant by *drift*. In this case a similar collection of features are analyzed to the one found in table 3. A treatment of these classification methods and supervision methods is found in Hu et al. (2007). A direct application of these tools to Wikipedia data would be to analyze the page visits[3] to certain topics so that one can discover topics that are interesting not in terms of absolute number of visitors, but are gaining or losing interest relative to themselves.

The question of governance is somewhat more general than the question of quality and is treated in section 3.4.

### 3.1.2 Evolution

Since the entire revision history is available for any article, comparisons among past versions of an article and its present state are an important source of data. Zeng et al. (2006) use statistical methods on this data to determine what they call "trust." As mentioned above, distinctions among "trust," "trustworthiness," "quality," and "reputation" are not made consistently in the literature. These authors do dig deeper into the definition of trust, by making the distinctions among

- Article trust: trustworthiness of a version (such as the current version) of an article;

- Fragment trust: trustworthiness of any fragment of text within an article; and

---

[3]Found at grok.se.

- Author trust: in particular, an author's trustworthiness across varying domains.

The authors focus on article trust, using a Dynamic Bayes Network to model the features that they extract from the article's revision history. In a radically simplified manner, they define the state of the DBN as a quadruple $\mathbf{X}_i = (t_{V_i}, t_{A_i}, i_i, d_i)$, where $t_{V_i}$ is the trust of version $i$, $t_{A_i}$ is the trust in the author of revision $i$, $i_i$ is the amount of inserted text into revision $i$ and $d_i$ is the amount of deleted text. Given a few independence assumptions and particular dependencies in the graph structure, the network's conditional densities can be fully characterized by $f(t_{V_0}|t_{A_0})$ and $f(t_{V_{i+1}}|t_{V_i}, t_{A_{i+1}}, i_i, d_i)$. To ease computation, Beta distributions are naturally selected for these conditional densities since their respective random variables' supports are $[0, 1]$. The parameters for the above Beta densities are selected using natural heuristics on the conditional terms. Furthermore, subjective Beta priors are selected for the distributions of $t_{A_i}$ depending on the Wikipedia-supported four levels of authorship: administrator, registered author, anonymous author, and blocked author.

The results of the model are compared to the Wikipedia-supported method of supervision by classifying articles as "featured", "normal", or "clean-up". The results of Zeng et al. (2006) show a signal, but one which is masked by so many heuristics that it is difficult to interpret or use for inference.

Thomas and Sheth (2007) measure the semantic stability of an article as a proxy for its quality. This proxy is justified using a two reasonable and insightful hypotheses.

> **Hypothesis 1** *A document can be seen as mature if, despite ongoing changes, it is semantically stable.*
>
> **Hypothesis 2** *A document is semantically stable if, after the kth revision, no significant changes have been made until the current nth revision, with $(n-k) > t$ being above a stability threshold t.*

To describe semantic difference, the authors acknowledge the lack of a "semantic oracle" $SO$, which is a function that maps to version of the article to $\{T, F\}$, depending on whether they have the same semantic[4] content. To remedy this, the authors approximate the semantic distance between two documents by considering each document as a $tf.idf$ vector, and measuring the cosine distance between two such vectors.

---

[4] Note that here, *semantic* does not have the same definition as in other contexts of this summary, especially regarding the Semantic Web.

$tf.idf$ stands for Term Frequency - Inverse Document Frequency, which is a common measure used in statistical natural language processing (see Manning and Schütze, 2002, p. 542). It is well-known that it provides a good estimate of the topic of an article relative to other articles, especially in the field of Information Retrieval. Once again, however, one may label $tf.idf$ as a feature of a page.

Looking at a plot of these edit distances over time will give any human eye a clear vision of this notion of semantic stability. The next natural step is to discover what generally happens around this convergence, how long it takes, and what other features of the article imply convergence at the same time. Since these features are likely to vary among articles, designing a learning machine which can adapt to the context of an article is the only way to account for how wildly different two articles' histories may be.

Unlike with automatic vandalism detection, training machines to do this work is valuable because it is beyond the breadth of any individual to actively retrace article histories, nor is it feasible to organize a group to do so. However, this learning is essentially unsupervised and so many heuristics get thrown into the design that usually such machines cannot learn anything beyond the heuristics that are used to design them. Essentially, they are not identifying a *natural* underlying process which a human reader could not discern, but rather change their parameters to line up with the features of the article that the designer considers important.

As another example, Thomas Wöhner and Ralf Peters (2009) identify that, given the nature of the wiki process, the *lifecycle* of a page is something common to all articles. That is, they invariably start from nearly nothing and evolve into a stable and hence reliable article. Identifying what stage its lifecycle an article currently finds itself gives an estimate of its reliability. The authors propose several features that can be measured from a revision and analyze the usefulness of these features. They do not go so far as to propose a learning machine with a heuristic notion of lifecycle as supervision, but that would be the next natural step in such research.

This paper does not propose that the solution is to design another learning machine but rather to propose methods of learning from the wiki process itself. That is, the wiki process, with its millions of editors, provides its own human-like intelligence,[5] which intelligence is so large that statistics and computers are necessary to leverage it.

---

[5]We say human-*like* because there is not one author, but many who play out a political interchange.

### 3.1.3 Stochastic Processes

Lerman and Hogg (2009) propose a significantly different way to view Wikipedia and its more general form, User-Contributory Web Sites. The authors propose that modeling an individual's behavior, when aggregated across a large-scale network, can accurately describe the activity across the entire web site. Modeling individuals is somewhat out of the scope of what is feasible within Wikipedia without drastic simplifications. Rather, the authors propose the modeling for the news aggregation site Digg, since an individual's actions can be completely characterized by his "Digg" or "Bury" votes on different stories along with timestamps. The case of Wikipedia involves vast amounts of unstructured text, although this type of stochastic modeling may one day be useful for it.

The authors choose a model for an individual and then simulate the behavior of the entire network by solving stochastic differential equations. The main assumption is that not unlike a Markov chain, in which the next state is probabilistically dependent on the current state, here the rate of change out of that state is dependent on the current state. Thus if there are $K$ states, and $n_k$ denotes the number of users in state $k$, writing $\vec{n} = (n_1, \ldots, n_K)$,

$$\frac{dn_k}{dt} = \sum_{j=1}^{K} n_j \cdot w_{jk}(\vec{n})$$

where $w_{jk}$ is a weight function that gives the transition rate. More complex equations including higher order derivatives can be proposed and solved.

It would be an astronomical task to find a differential equation that would model the activity on Wikipedia, but we can see here that the goal of discovering such a differential equation differs from the goals of the learning machines discussed above. Should a stochastic differential equation accurately reflect users' aggregate motions on Wikipedia, then one can remodel how the typical behavior of the user himself changes based on the user interface or software. Then one can use this information to try to optimize the user interface to arrive that the type of aggregate behavior one would like to see. This is very context-specific, but in the case of Wikipedia, goals could include routing knowledgeable people to underdeveloped articles or encouraging users to look at page histories or citations more often.

## 3.2 Trust

Adler and de Alfaro (2007) open up the problem of quality determination in Wikipedia by making distinctions among the terms "trust," "reliability," and "reputation."

They formally define the concepts so that they can be measured within Wikipedia.

The *trust* of a word block in an article in Wikipedia is measured on a finite numerical scale, for convenience $\{1, \ldots, 9\}$. It is a measure of how much the collection of users of Wikipedia (often called the "crowd") believe in the accuracy of the text in the word block. Thus it has two perspectives: first is how much the crowd *has trusted* the block of text and second how much the crowd *will trust* it. Trust in this sense becomes somewhat detached from what we'd call *accurate* or *correct* in science. It is rather a combination of consensus and Wikipedian self-governance. These topics are discussed in sections 4.1 and 3.4.

The trust calculation uses as its input the revision history of each article as well as information about the *reputation* of the contributing authors. This reputation is computed from revision histories as well. Intuitively, an author whose edits do not get removed by other high-reputation authors should have high reputation, whereas an author whose edits are quickly removed, especially by high-reputation authors, should have low reputation.

This seems to beget a chicken-and-egg problem but is quickly solved by assigning some optimized tuning parameter as a starting point for all authors.

However, what this means is that trust is calculated within Wikipedia and therefore supervised by the wiki system itself. Whereas the machine learning algorithms discussed above are supervised by the greatly reduced statistic of "featured article" status, or by their own heuristics, the content-driven reputation system developed here leverages the omnipresent supervision of the crowd.

The goal is therefore turned on its head. Rather than designing a machine to learn from the crowd, the crowd *is* the machine. Whereas a learning machine has perfect information (that is, data processing), but lacks intelligence beyond its design, the crowd has (political) intelligence, but lacks perfect information. Therefore the problem becomes, how can more relevant information be delivered to the crowd? Which statistics, when delivered to the crowd, make it work better, and which slow it down?

## 3.3 Semantic Extraction

The vision of the Semantic Web as proposed by Berners-Lee et al. (2001) has generated excitement in the computer science world but has stalled in its development. Discussion of the implications of a well-formed Semantic Web and the technology necessary to build it are beyond the scope of this summary, but motivation and most technical information can be found in Segaran et al. (2009).

The goal is to bootstrap semantic information using an autonomous self-supervised tool. Auer and Lehmann (2007) directly import semantic information from existing infoboxes in Wikipedia. Wu and Weld (2007) go one step further and attempt to extract semantic information from the body of the article using natural language learning machines. Wikipedia is the prime target for such efforts for a variety of reasons:

- Every important concept has a *Unique Resource Indicator* (URI) as indicated by the concept's URL.

- Homonyms that appear in wikilinks are thus disambiguated by the page they point to.

- Infoboxes provide ontologies of an object's key attributes.

- Lists and categories provide a simple type system and taxonomic hierarchy, respectively.

- Redirect pages can be used to induce synonyms.

- Disambiguation pages provide a list of candidate targets for homonym resolution, should the homonym not already be settled by a wikilink.

- Wikipedia is large enough to include the core concepts necessary for launching the Semantic Web, but not too large to process like the Web itself

The authors develop very sophisticated natural language processing techniques in Wu and Weld (2008). The goal in extracting such semantic information is to induce an artificial intelligence with all such data, treating metadata as data. The authors are working on ways to feed this intelligence back into the Wikipedia system. In particular, they created a user interface in which such extracted and synthesized information are fed back onto the Wikipedia page itself, for users to verify or refute (Hoffmann et al., 2009).

This sort of strategy is particularly fascinating because, as discussed above, these researchers are not attempting to build an artificial intelligence beyond the crowd. Rather, they are using a learning machine to produce results which are then *reabsorbed* into the wiki system. Such a system need not resort to heuristics to judge its efficacy. Rather, whether the results are absorbed by the crowd is an immediate and irrefutable proof of efficacy.

## 3.4 Governance and Society

Wikipedia can quickly become a source of controversy given differing opinions individuals have over what is considered "factual." As discussed above, a simple "majority rules" can stifle emerging theories or news. Furthermore, since editing is far from being distributed uniformly, any such voting or quasi-democratic system is prone to failure.

Rather, Wikipedians have instituted the *rule of law*. This paper does not propose to engage in political ideology but recognizes that the rule of law exists in all societies that are considered successful. Following Beschastnikh et al. (2008), we refer to this collection of laws as Wikipedia's *policy environment*. Each policy within the environment has its own Wikipedia page detailing the guidelines of the policy. Examples include WP:AGF, Assume Good Faith, and WP:EW, Edit Warring. These pages are developed and maintained by all Wikipedia users, not just administrators, and as such are subject to Wikipedia policy as well.

Wikipedia law is brought into play with *policy citations*, which appear in article discussion pages and wikilink to the policy under attack. Wikipedia law is qualitatively treated in Viégas et al. (2007), who claim that the law works due to the structure of the wiki system.

Beschastnikh et al. (2008), however, *quantitatively* attempt to prove the efficacy of Wikilaw by showing that unique users who cite policy for the first time account for over 10% of all unique users who city policy consistently over time, even with Wikipedia's exponential growth. This high percentage indicates that the governance structure is inclusionary, that it is not being co-opted by a select subset of users.

Beyond laws and policy, one soon realizes that Wikipedians participate in an online community, a society based on communication, respect, and a shared quest for the truth. Each active user does not nearly engage in the same collection of editing activities as everyone else, rather any Wikipedian carves out a role for himself with some degree of specialization. To quantify the level of specialization would be absurd, but Kriplean et al. (2008) develop a theoretical lens for uncovering valued work in Wikipedia through a content analysis of Wikipedia *barnstars*.

Barnstars are templates found in Wikipedia's meta namespace and are given by users to other users as gifts as a sign of appreciation for good work (see figure 1). There is no cost or limit to giving out barnstars beyond one's own reputation. The barnstars are prominently displayed on the user's user page (which, as a wiki page, anyone can edit), and reflect a level of pride and achievement on the part of the user. Furthermore, barnstars have been customized to reflect different types of work, such as the Photographer's Barnstar for those who improve Wikipedia with their photographic skills, or the Guidance Barnstar, for those who help others locate valuable resources, information, or assistance. Kriplean et al. (2008) point out that "the nuances behind giving a barnstar often reflect complex social dynamics, similar to gift-giving in any community."

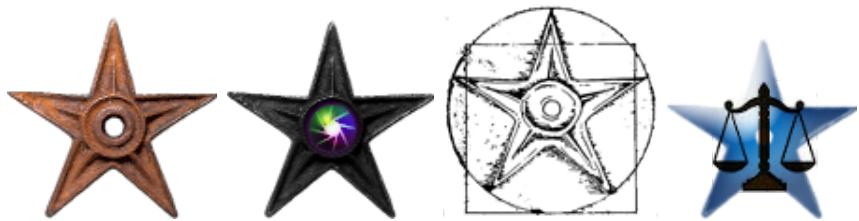

Figure 1: Some Wikipedia Barnstars.

The authors built a parser to extract barnstars and through iterated sampling develop a *codebook* of the varieties of Wikiwork that barnstars acknowledge. They are then able to develop some statistics that show which types of work are considered most important through this lens, by indicating the relative proportion of barnstars that are given out for that type of work.

Their results are highly nuanced, as expected, and expose the complex nature of building and maintaining Wikipedia. Indeed this encyclopedia is not just the result of a heap of edits, but rather a project and community that reflects the depth of the logic under which human society operates. Understanding this logic, that is, the logic of the crowd rather than an individual, will be essential for understanding and engineering artificial intelligence in the decades to come.

## 4   Philosophical Considerations

The existence of the internet, the World Wide Web, and now Wikipedia, throws a wrench into the systems of philosophers. As the world shifts toward more open-source software, more efficient information sources, and more free information, the usual models of free-market and command economies become even more theoretical. Wikipedia has its own socioeconomic structure, one that has only partially been seen before. Economists are confounded by an economy of free goods which nonetheless continues to produce.

Epistemologists are faced with an even larger task. Should Wikipedia "level off," or stop growing, does this represent the extent to human knowledge? Or must we then redefine human knowledge as the extent that knowledge is available to *everyone* and not just *anyone*?

## 4.1 Economic Implications

Wikipedia provides a means for distribution of the editing resource. This resource is scarce (that is, finite), along with page views, image uploads, and administrative action. It is well-documented that these resources are distributed according to a power law (Voss, 2005; Dondio and Barrett, 2007; Thomas and Sheth, 2007; Warren et al., 2008; Ortega et al., 2008). Estimated parameters vary from study to study but as a socioeconomic resource a power law distribution is expected.

The open-source software (OSS) movement, which really gained steam during the development of Linux, has always exercised centralized control on the development of code. A particular developer might choose to branch the code to work out his ideas, but that branch cannot be reintegrated without approval from the core developers. Olleros (2008) points out the distinction between Wikipedia and these OSS projects' decentralizing the quality control task.

The source of decentralizing quality control in Wikipedia is the decoupling of the results. Indeed computer programs must pass the so-called "silicon test," either it runs or it does not. But Wikipedia neither as a site nor a process is stopped by bad information. Olleros (2008) refers to this loose coupling as "natural firewalls," saying that "bad quality is always local and stays local."

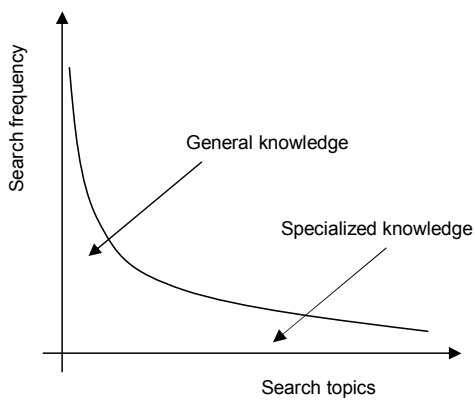

Figure 2: Distribution of search frequency to specialization of search topics. Adapted from Olleros (2008).

Using economic reasoning, Olleros (2008) goes further to say that Wikipedian quality control *must* be decentralized. In figure 2, the topics to the far right of the curve are highly specialized, and therefore rarely accessed. The left of the curve can be considered the topic space of general encyclopedias, whereas the tail end constitute the topic space of specialized encyclopedias. In the defunct Nupedia, the editorial team was tasked with monitoring this tail, an impossible burden given that specialization must continue to evolve as the encyclopedia grows beyond general knowledge and into specialized knowledge.

This notion of specialization diverges from easy quantification not only due to the *ad hoc* nature of measuring "specialized knowledge," but also that specialization implies a division of labor of Wikipedian tasks, as discussed in section 3.4. Similar dynamics evolve in the world's economy, where the collection of tasks produced by human labor cannot be quantified by level of specialized knowledge, but rather by level of specialized role, which depends on specialized knowledge among many other factors.

Olleros (2008) also discusses the most dramatic difference between the Wikipedian economy and a natural economy, that Wikipedians do not receive tangible benefits in exchange for their labor. "A job well done" and barnstars on one's user page does not feed his family or pay his rent. However, the *transaction cost* to doing anything on Wikipedia is negligible, especially given that one can make anonymous edits. The author claims that this anonymity played a critical role in Wikipedia's early growth. Anonymous users could contribute without risking their reputations, spending only their time.

Anonymity therefore produced a very low barrier to entry, and given at least a modicum of satisfaction that comes with contribution, the benefits to contribution outweighed the costs and Wikipedia grew. To explain such satisfaction is beyond the scope of this paper.

## 4.2 Epistemological Implications

Olleros (2008) goes on to claim that the notion of a "useful encyclopedia" must therefore be redefined given the new technology. By citing Christensen (1997), the author refers to Wikipedia as a "disruptive technology," one which underperforms along established dimensions of value but outperforms in new dimensions of value, thereby redefining what valuable technology is. It has done so along several dimensions: breadth, affordability, currency, internal and outgoing links, multiple languages, and openness to non-contractual contributions.

Furthermore, Wikipedia has opened up the critical dimension of accessibility and visibility. Fallis (2008) claims that this require clarification of our epistemic values. No longer is quality the sole metric with which to judge information. The question must be broken into three distinct philosophies: *epistemology of testimony, social epistemology,* and *epistemic value theory.*

An encyclopedia is a transmitter of knowledge from one group of people to another. Its purpose is to disseminate, not create, knowledge. It represents a *group testimony* presented to another social group. So whereas most of the philosophy of epistemology is concerned with the cognitive and perceptual processes within an individual, social epistemology is concerned with how social processes lead to the acquisition of knowledge. Therefore one must redefine what a *good* epistemic consequence is as related to a group.

When aggregating the need for knowledge across many individuals, the *principle of least effort* (Pareto, 1901) grants that dimensions beyond quality become relevant. The new dimensions of *power, speed,* and *fecundity* are evaluated. That is, we are concerned with *how much* knowledge can be acquired from an information source, *how fast* that knowledge can be acquired, and *how many* people can acquire that knowledge (Fallis, 2008).

There is therefore a trade-off between magnitude in the dimensions of quality and reliability and the the dimensions of power, speed, and fecundity. The power law depicted in figure 2 gives an abstraction of this trade-off. Given the high dimension and abstract nature of the features being measured, a two-dimensional representation can only be used to give the most basic intuition of this trade-off. The boundary, however, is observable in practice. Indeed throughout this survey many measures of quality and power have been discussed. But when abstracted to represent a trade-off between old and new epistemic values, Wikipedia can be used to define the epistemological frontier, the boundary beyond which humanity cannot produce as much accurate knowledge diffused to as many people.